# Self-Focused Electron Beams Produced by Pyroelectric Crystals on Heating or Cooling in Dilute Gases

J. D. Brownridge, State University of New York at Binghamton, P.O. Box 6016
Binghamton, New York 13902-6016 and
S. M. Shafroth, Physics and Astronomy Department, University of North Carolina at Chapel Hill 27599-3255

## ABSTRACT

Self-focusing, spatially stable, electron beams arising from heating up to $160^0$ C and cooling to room temperature pyroelectric $LiNbO_3$ crystals in dilute gases have been observed for the first time. The current (up to nanoamperes) and energies (up to 170 keV) of these electron beams attained maximum values and then decreased during the cooling phase of the thermal cycle. Cylindrical crystals produced spatially stable beams with typical focal lengths of 22 mm and 1mm spot sizes. Here we present photographic as well as electronic proof of their existence.

## INTRODUCTION

Although some of the remarkable electrical properties of pyroelectric crystals have been known since ancient Greek times [1], surprising new effects and practical applications are constantly being discovered [1-6]. There is a voluminous literature on electron emission by pyroelectric crystals. Reference 6, a comprehensive review article and (7-11) are representative but none of them suggests that electron beams are emitted by such crystals. Pyroelectric crystals become electrically charged on heating or cooling and can produce electric fields at their surfaces of up to $10^6$ V/cm. If the crystal is cut perpendicular to its three-fold rotationally-symmetric z axis, the surface called the -z base becomes positively charged on heating and then negatively charged on cooling. Much of the current industrial use of these crystals is in infrared detection, sensitive temperature change detectors and photonics, e.g., the detection of THz radiation [12] in $LiNbO_3$. This makes plausible the fact that the electron-beam phenomenon reported here has not been seen previously.

## DISCUSSION

This report describes the production of self-focusing electron beams arising from near pyroelectric crystal surfaces in dilute gases after the crystal has been heated and returned to room temperature. In one case a glass tube was used to house a 5 mm diam x 5 mm cylindrical $LiNbO_3$ crystal in an atmosphere of <10 mtorr of dry $N_2$. The crystal was heated to $115^0$ C and then allowed to cool to room temperature. Heat was supplied by passing ~100 mA through a wire-wound 60 ohm resistor which was epoxied to the crystal's + z base. After cooling to room temperature, spatially stable electron beams were produced. They were made visible by placing a ZnS screen at the focal length (17 mm) of the crystal as illustrated in Fig. 1(a), which shows a photograph of the ZnS screen; the cylindrical crystal and the 60 ohm wire-wound resistor used to heat the crystal. We studied the effect of pressure on the beam. Typical results at ~0.5 mtorr 1(b), ~3 mtorr 1(c) and ~8 mtorr 1(d) are shown in Fig 1(b), 1(c), 1(d) respectively. As the pressure increased, the beam spot became more diffuse and brighter. At ~8 mtorr the beam blew up and its intensity dropped to zero. The dynamical behavior of the electron beam with pressure suggests a gas-multiplication effect. All results reported on in this paper were repeated many times.

Next a 100 (m surface-barrier electron detector replaced the ZnS screen. The experimental arrangement is shown in Fig. 2. The crystal and heater assembly could be moved toward or away from the detector, which was useful for checking the focusing effect. The detector was covered by a lead screen with a 0.1mm slit so that the beam profile could be obtained by moving the slit-detector assembly at constant speed across the beam at different crystal-to-slit distances. Fig 3 shows results of scans at different crystal-to-slit distances. The best focus occurs when that distance is 17 mm. A further illustration of the focusing effect is seen in Fig. 4 where a different cylindrical crystal (4mm diam x 10mm $LiNbO_3$) and heater assembly was moved in and out relative to the detector slit. Here the focal length was 22 mm and the focusing effect was very dramatic. Once the crystal reached room temperature the beam energy decreased nearly exponentially with time (Fig. 5); faster with increasing pressure. The time-dependent electron-beam energy spectrum which was taken in "snapshot mode", i.e. over an interval of <60 s, typically, exhibited discrete peaks at integral multiples of the lowest energy peak. This indicated that multiple electron production[4] was occurring. In order to produce higher energy beams the crystal was heated from the + z base to $160^0$ C and allowed to cool to room temperature at <10 mtorr. It produced maximum energy electron beams of up to 145 keV when the temperature was ~$30^0$ C. That the electron beam energy was at least 80 keV (Au K edge) for a significant time was confirmed by irradiating gold and observing the K X-rays [2,3].

The focusing effect shows that the electric field lines near the crystal surface are slightly inclined toward the axis, which is reasonable since the field at the edges of the crystal is higher than at the center. If the crystal is subjected to a rapid temperature change the polarization will be manifested briefly by the presence of an electric field whose strength is proportional to the surface charge density, which in turn is a function of the change in temperature and depends on the ambient pressure. When the crystal is in a reduced pressure environment and is subjected to the same temperature change, the polarization change is manifested by an electric field that lasts much longer; at a pressure of about $10^{-6}$ torr, it takes more than 30 hours for the resulting surface charge to be neutralized, whereas at 1 atm neutralization is virtually instantaneous. The neutralization of the electric field produced by a change in polarization is due to bombardment of and attachment to the surface by positive ions. In conclusion, as long as the net surface charge is sufficiently negative, electrons are accelerated away from the crystal in a focused beam whose energy is constantly changing and after the crystal has reached room temperature, even before in some cases, the beam is spatially stable.

The above described results together with the recent discovery that at pressures in the $10^{-5}$ torr range, focused 100 keV positive-ion beams are generated on heating the crystal were reported[5] at the Beams Division of the Spring Meeting of the American Physical Society.

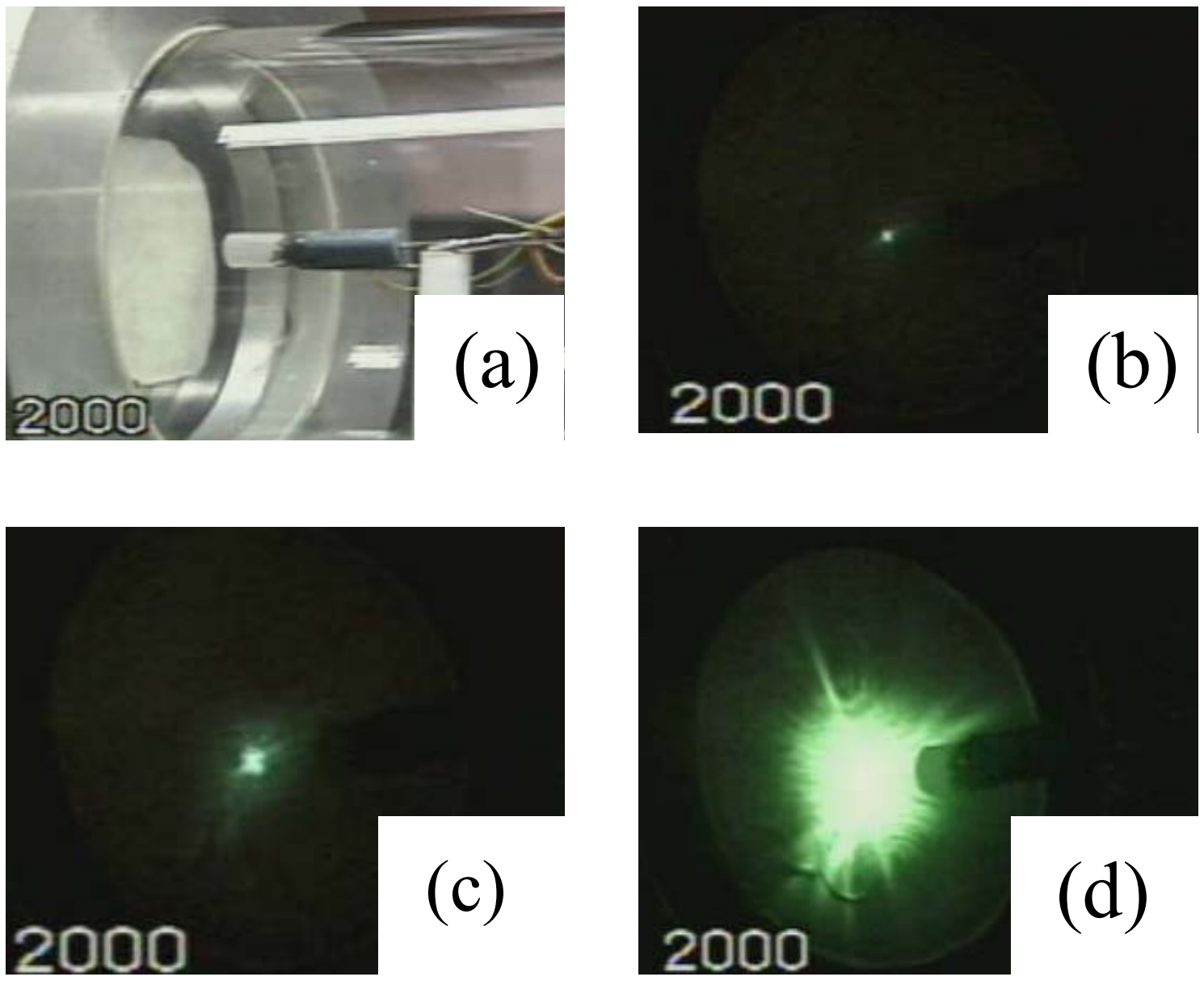

Fig. 1. (A) Photograph of the experimental arrangement with ZnS screen, crystal, and heater resistor, (B) Beam spot at ~0.5 mtorr, (C) ~3 mtorr, (D) ~8 mtorr, where the beam blows up.

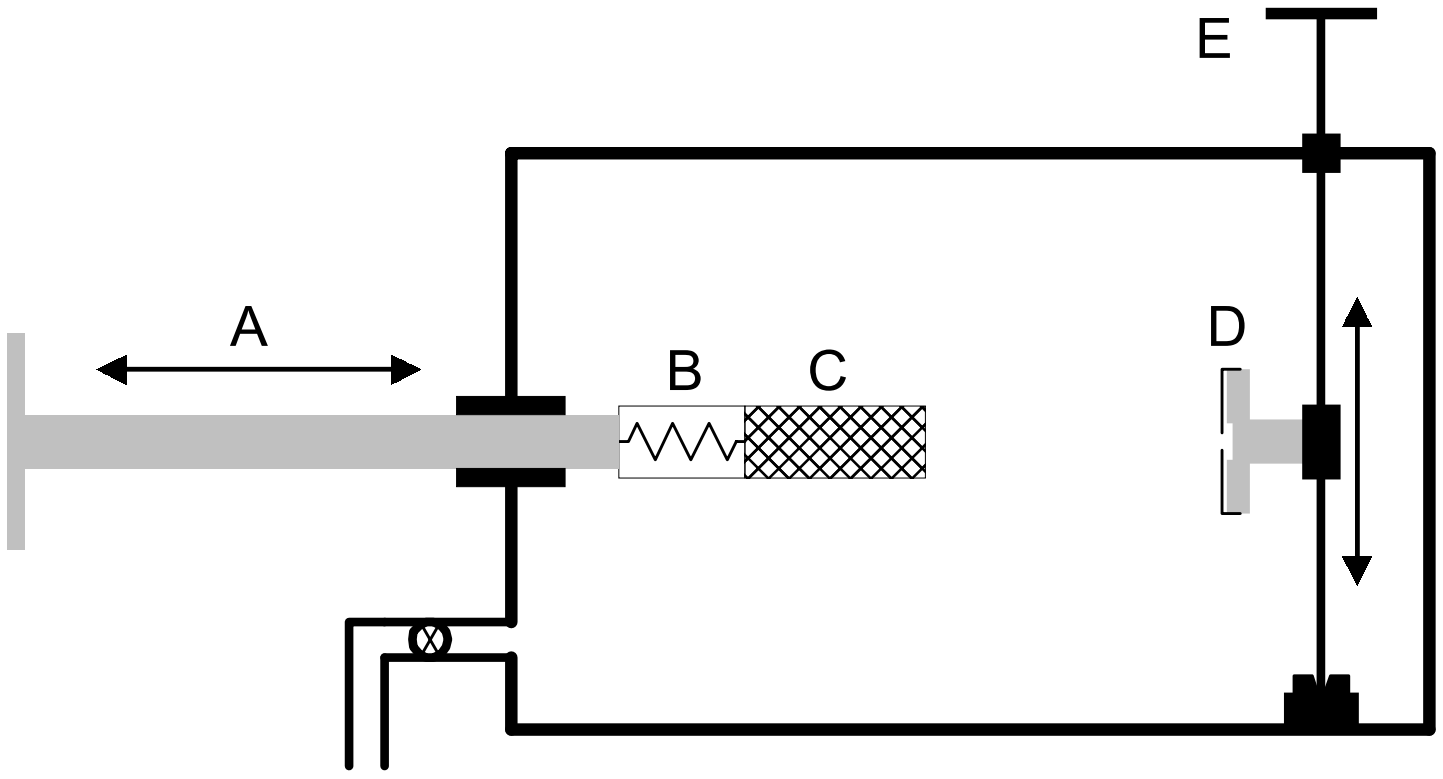

Fig.2 Experimental arrangement for direct electron detection, beam scanning and changing crystal to detector distance. (A) heater resistor and crystal mounting rod. (B) Heater resistor, (C) crystal, (D) slit and detector assembly, (E) Scanner motor.

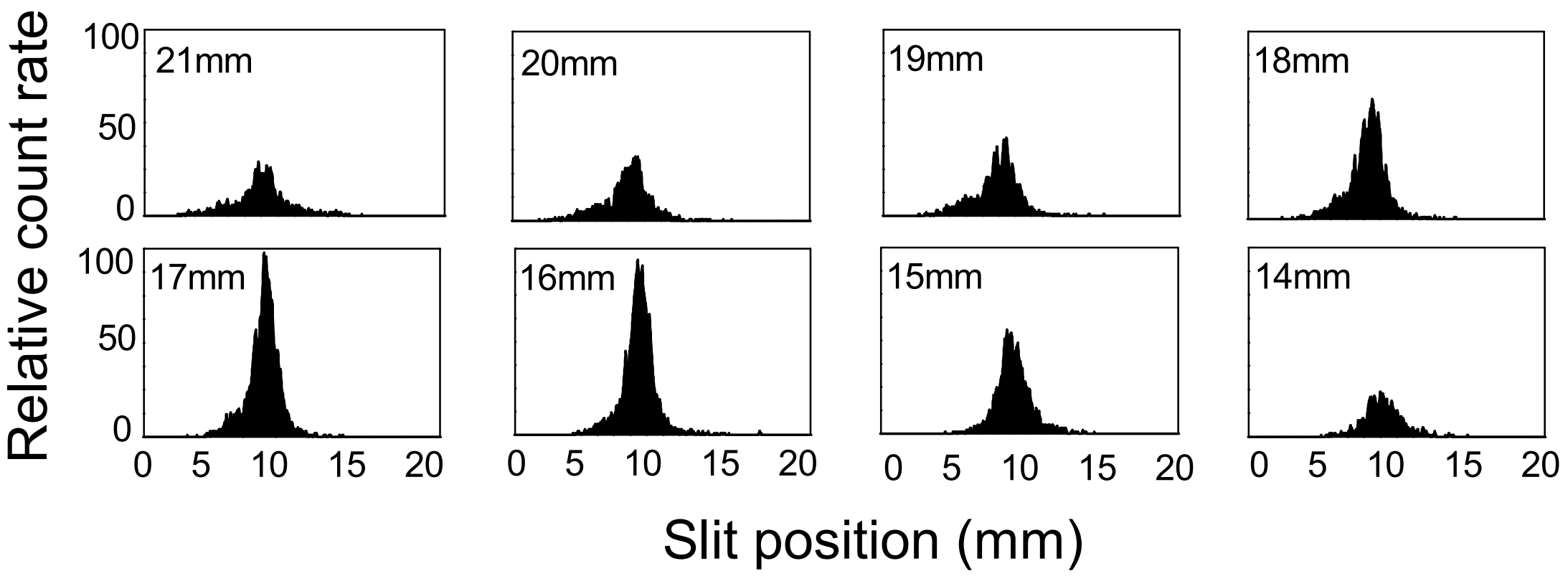

Fig. 3 Results of beam scans at different crystal-to-slit distances.

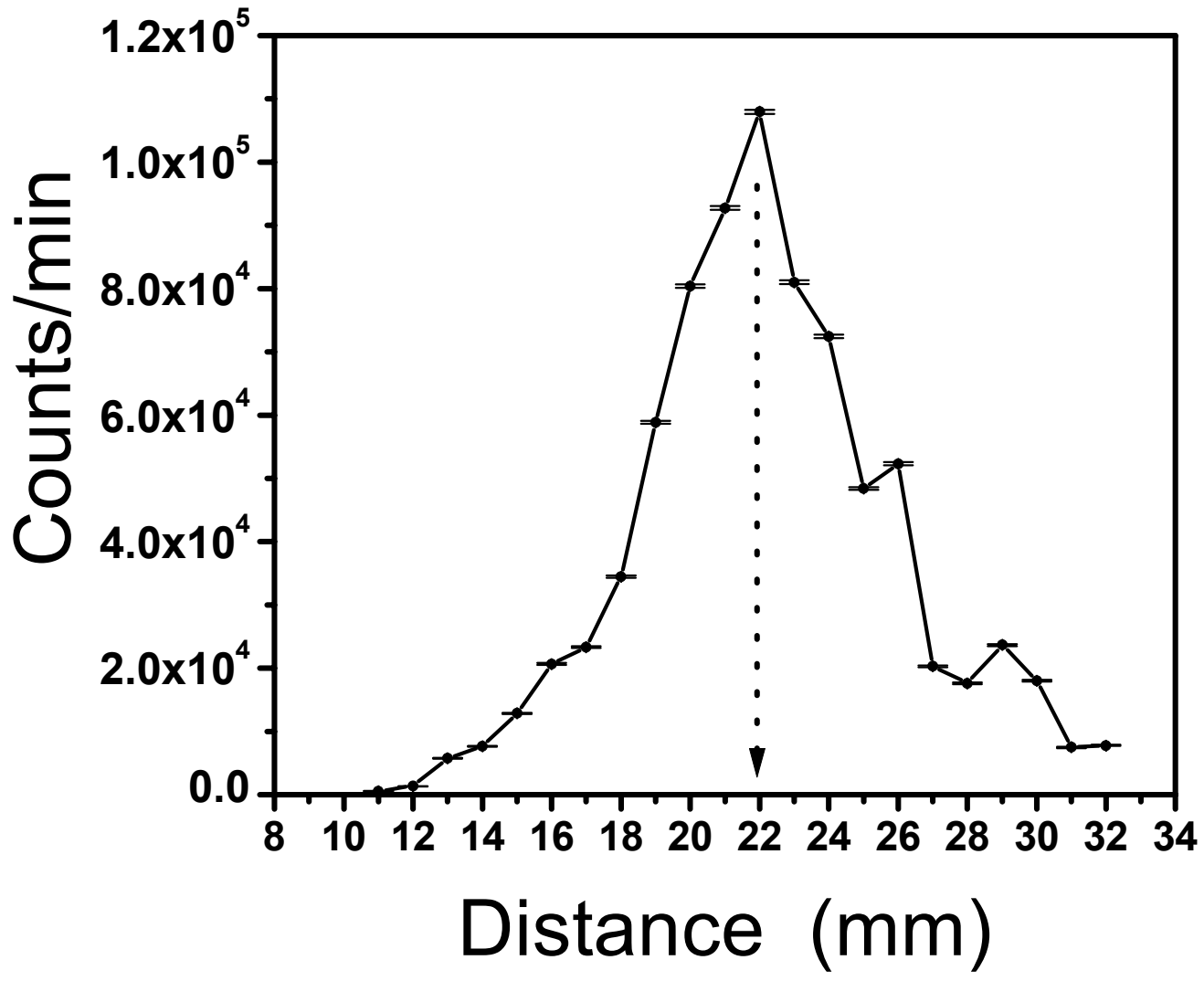

Fig. 4 Beam intensity as the crystal-heater unit is moved forward and back with respect to the detector-slit assembly. Here the focal length is 22 mm.

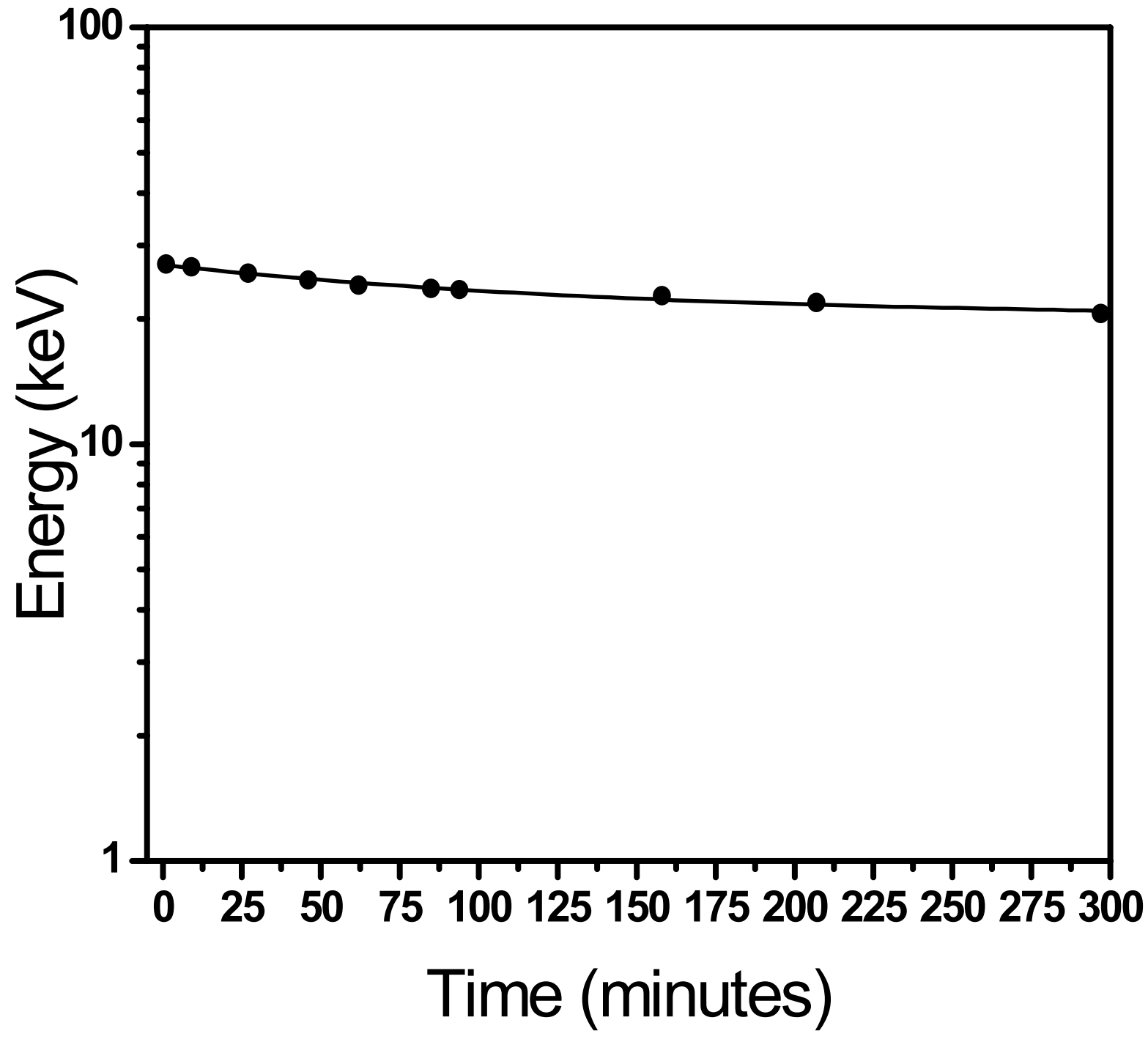

Fig. 5 Beam energy as a function of time after the crystal has reached room temperature. Note the nearly exponential decay of the beam energy.

jdbjdb@binghamton.edu